\RequirePackage{lineno}
\documentclass[aps,prd,twocolumn,superscriptaddress,letterpaper,floatfix,nofootinbib,showpacs,preprintnumbers]{revtex4}
\usepackage{epsfig,units,color,ulem}

\begin{document}
\preprint{FERMILAB-PUB-12-620-E}

\title{Comparisons of annual modulations in MINOS with the event rate modulation in CoGeNT}



\newcommand{\Berkeley}{Lawrence Berkeley National Laboratory, Berkeley, California, 94720 USA}
\newcommand{\Cambridge}{Cavendish Laboratory, University of Cambridge, Madingley Road, Cambridge CB3 0HE, United Kingdom}
\newcommand{\Cincinnati}{Department of Physics, University of Cincinnati, Cincinnati, Ohio 45221, USA}
\newcommand{\FNAL}{Fermi National Accelerator Laboratory, Batavia, Illinois 60510, USA}
\newcommand{\RAL}{Rutherford Appleton Laboratory, Science and Technologies Facilities Council, OX11 0QX, United Kingdom}
\newcommand{\UCL}{Department of Physics and Astronomy, University College London, Gower Street, London WC1E 6BT, United Kingdom}
\newcommand{\Caltech}{Lauritsen Laboratory, California Institute of Technology, Pasadena, California 91125, USA}
\newcommand{\Alabama}{Department of Physics and Astronomy, University of Alabama, Tuscaloosa, Alabama 35487, USA}
\newcommand{\ANL}{Argonne National Laboratory, Argonne, Illinois 60439, USA}
\newcommand{\Athens}{Department of Physics, University of Athens, GR-15771 Athens, Greece}
\newcommand{\NTUAthens}{Department of Physics, National Tech. University of Athens, GR-15780 Athens, Greece}
\newcommand{\Benedictine}{Physics Department, Benedictine University, Lisle, Illinois 60532, USA}
\newcommand{\BNL}{Brookhaven National Laboratory, Upton, New York 11973, USA}
\newcommand{\CdF}{APC -- Universit\'{e} Paris 7 Denis Diderot, 10, rue Alice Domon et L\'{e}onie Duquet, F-75205 Paris Cedex 13, France}
\newcommand{\Cleveland}{Cleveland Clinic, Cleveland, Ohio 44195, USA}
\newcommand{\Delhi}{Department of Physics \& Astrophysics, University of Delhi, Delhi 110007, India}
\newcommand{\GEHealth}{GE Healthcare, Florence South Carolina 29501, USA}
\newcommand{\Harvard}{Department of Physics, Harvard University, Cambridge, Massachusetts 02138, USA}
\newcommand{\HolyCross}{Holy Cross College, Notre Dame, Indiana 46556, USA}
\newcommand{\Houston}{Department of Physics, University of Houston, Houston, Texas 77204, USA}
\newcommand{\IIT}{Department of Physics, Illinois Institute of Technology, Chicago, Illinois 60616, USA}
\newcommand{\Iowa}{Department of Physics and Astronomy, Iowa State University, Ames, Iowa 50011 USA}
\newcommand{\Indiana}{Indiana University, Bloomington, Indiana 47405, USA}
\newcommand{\ITEP}{High Energy Experimental Physics Department, ITEP, B. Cheremushkinskaya, 25, 117218 Moscow, Russia}
\newcommand{\JMU}{Physics Department, James Madison University, Harrisonburg, Virginia 22807, USA}
\newcommand{\LASL}{Nuclear Nonproliferation Division, Threat Reduction Directorate, Los Alamos National Laboratory, Los Alamos, New Mexico 87545, USA}
\newcommand{\Lebedev}{Nuclear Physics Department, Lebedev Physical Institute, Leninsky Prospect 53, 119991 Moscow, Russia}
\newcommand{\LLL}{Lawrence Livermore National Laboratory, Livermore, California 94550, USA}
\newcommand{\LosAlamos}{Los Alamos National Laboratory, Los Alamos, New Mexico 87545, USA}
\newcommand{\Manchester}{School of Physics and Astronomy, University of Manchester, Oxford Road, Manchester M13 9PL, United Kingdom}
\newcommand{\MIT}{Lincoln Laboratory, Massachusetts Institute of Technology, Lexington, Massachusetts 02420, USA}
\newcommand{\Minnesota}{University of Minnesota, Minneapolis, Minnesota 55455, USA}
\newcommand{\Crookston}{Math, Science and Technology Department, University of Minnesota -- Crookston, Crookston, Minnesota 56716, USA}
\newcommand{\Duluth}{Department of Physics, University of Minnesota Duluth, Duluth, Minnesota 55812, USA}
\newcommand{\Ohio}{Center for Cosmology and Astro Particle Physics, Ohio State University, Columbus, Ohio 43210 USA}
\newcommand{\Otterbein}{Otterbein University, Westerville, Ohio 43081, USA}
\newcommand{\Oxford}{Subdepartment of Particle Physics, University of Oxford, Oxford OX1 3RH, United Kingdom}
\newcommand{\PennState}{Department of Physics, Pennsylvania State University, State College, Pennsylvania 16802, USA}
\newcommand{\PennU}{Department of Physics and Astronomy, University of Pennsylvania, Philadelphia, Pennsylvania 19104, USA}
\newcommand{\Pittsburgh}{Department of Physics and Astronomy, University of Pittsburgh, Pittsburgh, Pennsylvania 15260, USA}
\newcommand{\IHEP}{Institute for High Energy Physics, Protvino, Moscow Region RU-140284, Russia}
\newcommand{\Rochester}{Department of Physics and Astronomy, University of Rochester, New York 14627 USA}
\newcommand{\RoyalH}{Physics Department, Royal Holloway, University of London, Egham, Surrey, TW20 0EX, United Kingdom}
\newcommand{\Carolina}{Department of Physics and Astronomy, University of South Carolina, Columbia, South Carolina 29208, USA}
\newcommand{\SLAC}{Stanford Linear Accelerator Center, Stanford, California 94309, USA}
\newcommand{\Stanford}{Department of Physics, Stanford University, Stanford, California 94305, USA}
\newcommand{\StJohnFisher}{Physics Department, St. John Fisher College, Rochester, New York 14618 USA}
\newcommand{\Sussex}{Department of Physics and Astronomy, University of Sussex, Falmer, Brighton BN1 9QH, United Kingdom}
\newcommand{\TexasAM}{Physics Department, Texas A\&M University, College Station, Texas 77843, USA}
\newcommand{\Texas}{Department of Physics, University of Texas at Austin, 1 University Station C1600, Austin, Texas 78712, USA}
\newcommand{\TechX}{Tech-X Corporation, Boulder, Colorado 80303, USA}
\newcommand{\Tufts}{Physics Department, Tufts University, Medford, Massachusetts 02155, USA}
\newcommand{\UNICAMP}{Universidade Estadual de Campinas, IFGW-UNICAMP, CP 6165, 13083-970, Campinas, SP, Brazil}
\newcommand{\UFG}{Instituto de F\'{i}sica, Universidade Federal de Goi\'{a}s, CP 131, 74001-970, Goi\^{a}nia, GO, Brazil}
\newcommand{\USP}{Instituto de F\'{i}sica, Universidade de S\~{a}o Paulo,  CP 66318, 05315-970, S\~{a}o Paulo, SP, Brazil}
\newcommand{\Warsaw}{Department of Physics, University of Warsaw, Ho\.{z}a 69, PL-00-681 Warsaw, Poland}
\newcommand{\Washington}{Physics Department, Western Washington University, Bellingham, Washington 98225, USA}
\newcommand{\WandM}{Department of Physics, College of William \& Mary, Williamsburg, Virginia 23187, USA}
\newcommand{\Wisconsin}{Physics Department, University of Wisconsin, Madison, Wisconsin 53706, USA}
\newcommand{\deceased}{Deceased.}

\affiliation{\ANL}
\affiliation{\Athens}
\affiliation{\BNL}
\affiliation{\Caltech}
\affiliation{\Cambridge}
\affiliation{\UNICAMP}
\affiliation{\Cincinnati}
\affiliation{\FNAL}
\affiliation{\UFG}
\affiliation{\Harvard}
\affiliation{\HolyCross}
\affiliation{\Houston}
\affiliation{\IIT}
\affiliation{\Indiana}
\affiliation{\Iowa}
\affiliation{\UCL}
\affiliation{\Manchester}
\affiliation{\Minnesota}
\affiliation{\Duluth}
\affiliation{\Otterbein}
\affiliation{\Oxford}
\affiliation{\Pittsburgh}
\affiliation{\RAL}
\affiliation{\USP}
\affiliation{\Carolina}
\affiliation{\Stanford}
\affiliation{\Sussex}
\affiliation{\TexasAM}
\affiliation{\Texas}
\affiliation{\Tufts}
\affiliation{\Warsaw}
\affiliation{\WandM}

\author{P.~Adamson}
\affiliation{\FNAL}


\author{I.~Anghel}
\affiliation{\ANL}
\affiliation{\Iowa}










\author{G.~Barr}
\affiliation{\Oxford}









\author{M.~Bishai}
\affiliation{\BNL}

\author{A.~Blake}
\affiliation{\Cambridge}


\author{G.~J.~Bock}
\affiliation{\FNAL}


\author{D.~Bogert}
\affiliation{\FNAL}




\author{S.~V.~Cao}
\affiliation{\Texas}




\author{S.~Childress}
\affiliation{\FNAL}



\author{J.~A.~B.~Coelho}
\affiliation{\Tufts}
\affiliation{\UNICAMP}


% ...

\collaboration{The MINOS Collaboration}
\noaffiliation



\date{\today}

\begin{abstract}
 The CoGeNT collaboration has recently published results from a fifteen month data set which indicate an annual modulation in the event rate similar to what is expected from weakly interacting massive particle interactions. It has been suggested that the CoGeNT modulation may actually be caused by other annually modulating phenomena, specifically the flux of atmospheric muons underground or the radon level in the laboratory. We have compared the phase of the CoGeNT data modulation to that of the concurrent atmospheric muon and radon data collected by the MINOS experiment which occupies an adjacent experimental hall in the Soudan Underground Laboratory. The results presented are obtained by performing a shape-free $\chi^{2}$ data-to-data comparison and from a simultaneous fit of the MINOS and CoGeNT data to phase-shifted sinusoidal functions.  Both tests indicate that the phase of the CoGeNT modulation is inconsistent with the phases of the MINOS muon and radon modulations at the \unit[3.0]{$\sigma$} level.\\

\end{abstract}


\maketitle

\section{Introduction}

Numerous astrophysical observations strongly support the existence in our galaxy of a cold dark matter halo, that may consist of Weakly Interacting Massive Particles~(WIMPs)~\cite{Feng:2010gw,Bertone:2004pz}.  The principal search mode of direct WIMP detection is the identification of an $\mathcal{O}$(keV) nuclear recoil produced by WIMP-nucleus elastic scattering. Since the speed of the Earth relative to the dark matter halo varies depending on the Earth's velocity with respect to the Sun, the dark matter detection rate is expected to demonstrate a annual modulation.  This modulation is expected to be at a maximum~(minimum) on June~2~(Dec.~2) with an amplitude between a few and \unit[20]{\%}, assuming the standard halo model~\cite{Drukier:1986tm,Freese:1987wu,Frandsen:2011gi}.  The CoGeNT~\cite{Aalseth:2011wp,Aalseth:2010vx}, DAMA/LIBRA~\cite{Bernabei:2010mq} and CRESST-II~\cite{Angloher:2011uu} collaborations have all reported an excess of events above all known backgrounds.  The CoGeNT~\cite{Aalseth:2011wp} and DAMA/LIBRA~\cite{Bernabei:2010mq,Bernabei:2008yh} collaborations have also claimed evidence for annual modulations in their event rates at \unit[2.8]{$\sigma$} and \unit[8.9]{$\sigma$} respectively.  Fits to the available data favor a light WIMP with mass \unit[$\sim$10]{GeV/c$^{2}$} and spin-independent cross-section \unit[10$^{-41}$--10$^{-39}$]{cm$^{2}$}~\cite{Hooper:2012ft,Kelso:2011gd,Kopp:2011yr}.\\

The null observations by CDMS-II~\cite{Ahmed:2010hw,Ahmed:2010wy,Ahmed:2012vq}, XENON100~\cite{Angle:2011th,Aprile:2011hi,Aprile:2012hi} and EDELWEISS~\cite{Armengaud:2012kd} exclude much of the allowed WIMP signal regions mentioned above~\cite{HerreroGarcia:2012fu}.  The tension between these exclusion limits and the positive observations can be significantly reduced, but not removed, when taking into account experimental~\cite{Collar:2012ed,Collar:2011wq} and astrophysical uncertainties~\cite{Kopp:2011yr,Schwetz:2011xm,Farina:2011pw,Kelso:2011gd,Foot:2010rj,HerreroGarcia:2011aa,Frandsen:2011gi}.  This tension has led to suggestions that the CoGeNT and DAMA/LIBRA modulations are due to conventional annual phenomena~\cite{Ralston:2010bd,Blum:2011jf}.  The atmospheric muon rate and the radon level in the underground experimental hall modulate annually. Signals that can simulate dark matter interactions may be produced by ($\alpha$,n) reactions from radon decay in the active volume or by nuclear recoils from spallation neutrons originating from atmospheric muon interactions.  The CoGeNT collaboration has stated that contamination from these backgrounds is small compared to the observed signal~\cite{Aalseth:2012if,Aalseth:2011wp}. The MINOS experiment monitors both of these quantities in an adjacent experimental hall to that of the CoGeNT experiment in the Soudan Underground Laboratory.  In this paper we compare the modulations of the CoGeNT event rate data to that of the atmospheric muon rate and radon level data collected at the same time by the MINOS experiment.\\

The annual modulation of the muon flux deep underground has been observed by many different experiments~\cite{Ambrosio:1997tc,Bouchta:1999,Bellini:2012te,Solvi:2009,Desiati:2011,Adamson:2009zf}.  The similarities of the amplitudes and phases of the modulations observed in the LVD muon~\cite{Solvi:2009} and DAMA/LIBRA data sets motivated the hypothesis that modulation in the latter may be muon-induced.  It has been suggested that spallation neutrons or long-lived activated isotopes produced by these muons may be responsible for the DAMA/LIBRA modulation~\cite{Blum:2011jf,Ralston:2010bd}.  This now seems unlikely as recent detailed comparisons of the DAMA/LIBRA modulation to that of the muon fluxes measured by LVD~\cite{Solvi:2009}, Borexino~\cite{Bellini:2012te} and MACRO~\cite{Ambrosio:1997tc}, all in the Gran Sasso National Laboratory~(LNGS), have shown that the phases of the two modulations differ significantly~\cite{Bernabei:2012wp,FernandezMartinez:2012wd,Chang:2011eb}.  This conclusion does not preclude the possibility that the CoGeNT modulation, or a significant fraction thereof, is due to muon related processes.\\

The phase of the modulation of the muon flux can vary substantially depending on geographic location and calendar year since the flux is strongly correlated with the effective atmospheric temperature~\cite{Ambrosio:1997tc,Bouchta:1999,Bellini:2012te,Solvi:2009,Desiati:2011,Adamson:2009zf}.  Therefore, to be able to reject with high confidence the muon hypothesis as the source of the CoGeNT modulation, the muon data must be collected concurrently with the CoGeNT data and in close proximity to the CoGeNT detector.  The muon data collected by the MINOS experiment fulfill these criteria.   Similarly to the DAMA/LIBRA muon studies~\cite{Bernabei:2012wp}, we compare the phase of the observed MINOS muon modulation to that of the CoGeNT data modulation. Comparisons of the CoGeNT data to non-concurrent MINOS muon data~\cite{Adamson:2009zf}, and indirectly to effective temperature variations, have been presented in Ref.~\cite{Chang:2011eb} and indicate that the data sets are not correlated.\\

We note that the 16.6\% amplitude of the CoGeNT event rate modulation~\cite{Aalseth:2011wp} is significantly larger than the $\sim$2\% amplitude of the MINOS muon rate modulation~\cite{Adamson:2009zf}. This difference suggests that the muon temporal variation cannot fully account for the observed CoGeNT modulation. In this paper we examine the relative phases of the two modulations which provides an independent test of the potential correlation between the CoGeNT and MINOS muon data sets.\\

The radon level in the Soudan Underground Laboratory is at a maximum~(minimum) in the summer~(winter) months due to the pressure gradients created by the relative temperature differences between the air in the laboratory and that on the surface~\cite{Goodman:1999}.  In the MINOS cavern we have observed that the radon concentration varies by a factor of six over the year, corresponding to a modulation amplitude of $\sim$60\%.  A large modulation amplitude could therefore be introduced into the CoGeNT data by even a small amount of contamination from this background.\\

The radon progeny also modulate with a one year period $T$, but do so with a delayed phase and reduced amplitude. The decays between $^{222}$Rn and $^{210}$Pb occur very quickly~($\sim$minutes) and therefore have negligible impact on either the phase or the amplitude. Since $^{210}$Pb has a half-life of \unit[22]{years}, its decay and the decays of its progenies will not contribute to the modulation.\\

The following Section of the paper discusses the selected experimental data sets.  In Section~\ref{sec:TheDifferentTests} we present the best fit modulation parameters determined for each of these data sets.  We then describe the measurements of the phase differences between the CoGeNT and MINOS muon and radon data sets obtained from a simultaneous fit of the data to phase-shifted sinusoidal functions, a shape-free $\chi^{2}$ data-to-data comparison and a bin by bin correlation test.  Section \ref{sec:Conclusion} summarizes our conclusions.\\
\section{The Selected Data}
\label{sec:TheData}
The CoGeNT dark matter experiment~\cite{Barbeau:2007qi,Aalseth:2012if} and the Far Detector of the MINOS long baseline neutrino experiment~\cite{Michael:2008bc} are located \unit[705]{m} underground in two different caverns of the Soudan Underground Laboratory.  The MINOS cavern, which houses the MINOS detector, is \unit[82]{m} long, \unit[15]{m} wide and \unit[13]{m} high and is oriented along the direction of the NuMI neutrino beam~\cite{Crane:1995ky}. The CoGeNT and CDMS-II dark matter experiments are located in the Soudan 2 cavern which is similar in shape to the MINOS cavern but is \unit[70]{m} long and is oriented north-south. The two experimental caverns are connected by an east-west passage on their north side and are served by a common ventilation system which replaces the lab air several times per hour.\\
\subsection{The CoGeNT Data}
\label{sec:CoGeNTData}
CoGeNT is an experiment for direct detection of dark matter which employs a \unit[0.44]{kg} p-type point contact germanium detector~\cite{Aalseth:2008rx,Aalseth:2010vx,Aalseth:2011wp}.  The CoGeNT collaboration has published its results using data collected over a period of 458 days between Dec.~4,~2009 and Mar.~6,~2011 with a total of 442 live days~\cite{Aalseth:2011wp}.  The data were presented in fifteen 30-day intervals and one 8-day interval, then fit to a modulation hypothesis of the form:
\begin{equation}
R=R_{0}\left(1+A\cdot\textrm{cos}\left [ \frac{2\pi}{T}(t-t_{0}) \right ] \right ), 
\label{eq:cosine}
\end{equation}
where $R_{0}$ is the mean rate, $A$ is the modulation amplitude and $T$ is the period.  The time $t$ is the number of days since Jan.~1, 2010.  The phase $t_{0}$ is the day at which the signal is at a maximum.  The published CoGeNT best fit results are given in the last line of Table~\ref{tab:FitResults}.  The modulation hypothesis is preferred over the null hypothesis at \unit[2.8]{$\sigma$}.  The CoGeNT collaboration has released the background-subtracted data set used in this analysis to the public.  The results of our $\chi^{2}$ fit of the CoGeNT data to Eq.~(\ref{eq:cosine}), discussed further in Sec.~\ref{sec:CosineTest}, are in good agreement with the published results~\cite{Aalseth:2011wp}.\\
\subsection{The MINOS Data}
The MINOS Far Detector has been collecting atmospheric muon data since August 2003~\cite{Michael:2008bc,Adamson:2007ww}.  The experiment also records the radon level in the laboratory air.  The MINOS muon and radon data used in this analysis were collected between June~4,~2009 and Sept.~6,~2011.  This collection window is 12 months longer than the CoGeNT run period, from Dec.~4,~2009 to March~6,~2011, allowing the data-to-data comparisons described in Sections \ref{sec:ShapeFreeTest} and \ref{sec:CorrelationTest}.\\

The event selection and data quality requirements used in this analysis are identical to those in the previous study of seasonal muon intensity variation at the MINOS Far Detector, with the additional requirement that the reconstructed muon track be downward going. Restricting the data set to contain only days with greater than \unit[10,000]{s} of live time yields a total of 738~good days of atmospheric muon data. These good days include \unit[449]{days} which occurred between Dec.~4,~2009 and March~6,~2011 inclusive.\\
 
The radon level in the MINOS cavern air, inferred from counting the number of alpha decays, is measured every hour by a Model 1027 Sun Nuclear Corporation radon monitor~\cite{radon-monitor}.  A daily measure of the radon level is determined by averaging the 24 measurements taken throughout the day.  The standard deviation of these measurements, $\sigma$, is taken to be the error on the daily radon measurement. While larger than the standard error on the mean value, $\sigma/\sqrt{24}$, this choice is more consistent with the published accuracy of the radon monitor~\cite{radon-monitor}. There are 786 good days during which the radon monitor operated continuously throughout the day. These good days include \unit[458]{days}  which occurred between Dec.~4,~2009 and March~6,~2011 inclusive. The radon monitor was moved to different locations in the Soudan Underground Laboratory and cross calibrated with other detectors running simultaneously.  This demonstrated that the radon level does not vary spatially in the laboratory to within the resolution of the monitor.  Thus the radon levels measured in the MINOS cavern can be used to evaluate whether the CoGeNT data are correlated with the radon level in the Soudan cavern.\\
\begin{figure}[thb]
  \begin{center}
    \includegraphics[width=0.5\textwidth]{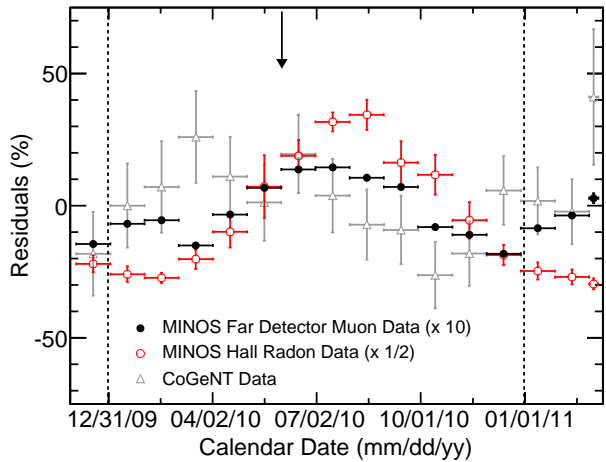}
  \end{center}
  \caption{The residuals of the MINOS Far Detector muon rates, radon levels and CoGeNT event rates as a function of time. The MINOS muon and radon data have been scaled by factors of 10 and one-half respectively to fit on the same graph and, for this figure, use the same binning as the CoGeNT results.  The vertical dashed lines indicate the start of a new calendar year. The arrow marks the date where a dark matter signal is expected to peak.}
  \label{fig:AllData}
\end{figure}

The MINOS muon rate and radon level residuals, and the CoGeNT event rate residuals, are plotted as a function of time in Fig.~\ref{fig:AllData}.  The CoGeNT event rate residuals are calculated with respect to a mean rate of \unit[97.7]{events/30~days}.  The MINOS muon rate residuals are calculated with respect to a mean rate of \unit[(0.4431 $\pm$ 0.0001)]{Hz}.  The MINOS radon level residuals are calculated with respect to a mean level \unit[(11.94 $\pm$ 0.11)]{pCi/l}.  All three data sets possess clear modulation signatures. In the following section we quantify any potential correlations between these modulations.\\
\section{Modulation Comparisons}
\label{sec:TheDifferentTests}

If the CoGeNT modulation is caused by either the muon or radon backgrounds then it should modulate with the same shape as those backgrounds. Therefore, if the phase of the CoGeNT modulation is significantly different than that of the MINOS muon or radon data we can infer that they are likely not causally related. \\

The most common approach in the literature to evaluating potential correlations, and discussed here in Sec.~\ref{sec:CosineTest}, is to fit the data to Eq.~(\ref{eq:cosine}) and compare the phases and periods of the best fits.  The CoGeNT and DAMA/LIBRA modulations are a good fit to a cosine function. This is the expected signature for an isothermal dark matter halo.  The true form of the modulation may be more complex as it is dependent on assumptions made regarding the velocity distribution of the dark matter particles in the halo~\cite{Freese:2012xd,Chang:2011eb}. The muon modulation is not fit well by a cosine function~\cite{FernandezMartinez:2012wd,Chang:2011eb}. The muon and radon modulations are correlated with atmospheric temperatures. Therefore, their modulations are cyclical but not necessarily sinusoidal. Imposing such constraints onto the data may bias the results of the cosine based fit comparison. We address this concern in Sections \ref{sec:ShapeFreeTest} and \ref{sec:CorrelationTest} by performing shape-free data-to-data comparisons that allow us to evaluate the phase differences and potential correlations regardless of the underlying functional forms of the modulations.\\

\begin{table*}[Htb]
\begin{tabular}{ccccccc} \hline
Data                & $\chi^{2}$/N.d.o.f.    & Mean Rate          & Amplitude     & Period       &Phase          & Date of          \\
                    &                   & [$R_{0}$] &  [$A$,\%]     & [$T$,days]     & [$t_{0}$,days]& Maximum          \\ \hline
\multicolumn{7}{c}{Best fit modulation parameters assuming a fixed period of \unit[365.25]{days}.}\\\hline
Muon                &1909~/~(449-3)     &  \unit[(0.4428 $\pm$ 0.0001)]{Hz}  & 1.25 $\pm$ 0.03 & 365.25          &182.8 $\pm$ 1.7& July 1   \\ 
Radon               &176~/~(458-3)      &  \unit[(11.9 $\pm$ 0.1)]{pCi/l}       & 57.7 $\pm$ 0.9  & 365.25          &215.0 $\pm$ 1.1& Aug. 3   \\ 
CoGeNT~(Our Fit)    &6.6~/~(16-3)       &  \unit[(97.9 $\pm$ 3.6)]{counts/30~days}       & 16.9 $\pm$ 5.4  & 365.25          &108.4 $\pm$ 16.9& Apr. 18\\ \hline 
\multicolumn{7}{c}{Best fit modulation parameters without a fixed period assumption.}\\\hline
Muon                &1788~/~(449-4)     &  \unit[(0.4431 $\pm$ 0.0001)]{Hz}  & 1.37 $\pm$ 0.04  &317.2 $\pm$ 3.2 &187.3 $\pm$ 1.4& July 6   \\ 
Radon               &176~/~(458-4)      &  \unit[(12.0 $\pm$ 0.1)]{pCi/l}       & 57.7 $\pm$ 0.9  &367.4 $\pm$ 3.5  &215.2 $\pm$ 1.1& Aug. 3   \\ 
CoGeNT~(Our Fit)    &6.4~/~(16-4)       &  \unit[(97.7 $\pm$ 3.6)]{counts/30~days}      & 16.7 $\pm$ 5.4  &348 $\pm$ 42     &113.7 $\pm$ 17.9& Apr. 23 \\ \hline
\multicolumn{7}{c}{Published CoGeNT modulation parameters~\cite{Aalseth:2011wp}.}\\\hline
CoGeNT              &7.8~/~(16-4)       &       N/A              & 16.6 $\pm$ 3.8  &347 $\pm$ 29     &$\sim$115 $\pm$ 12 & Apr. 25\\ \hline 
\end{tabular} 
\caption{The best fit results produced by fitting the MINOS muon rate, radon level and CoGeNT event rate data to Eq.~(\ref{eq:cosine}). The fits reported in the first three rows of the table have been performed with the period fixed to \unit[1]{year}~(\unit[365.25]{days}). The last column gives the dates in 2010 at which the fits to the data are at a maximum. }
\label{tab:FitResults}
\end{table*}
 \subsection{Cosine $\chi^{2}$ Test}
\label{sec:CosineTest}
\begin{figure}
  \begin{center}
    \includegraphics[width=0.5\textwidth]{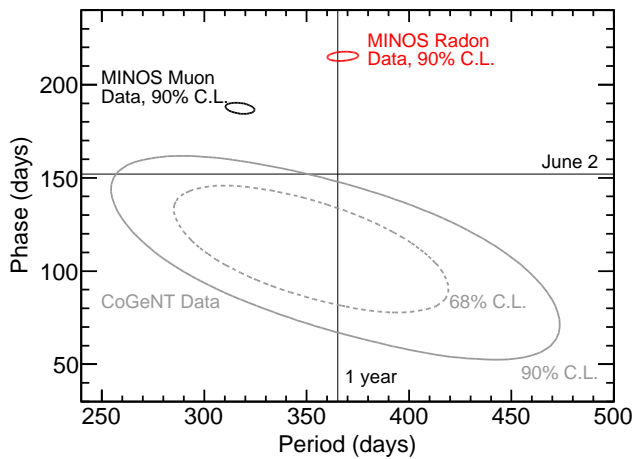}
  \end{center}
  \caption{Confidence limit contours for the period and phase as determined by fitting the CoGeNT event rate and MINOS Far Detector muon rate and radon level data to Eq.~(\ref{eq:cosine}). The best fit values are given in Table \ref{tab:FitResults}. The horizontal and vertical black lines mark the expected period and phase for a dark matter signal. }
  \label{fig:PhasePeriodFits}
\end{figure}

The nominal modulation parameters for the CoGeNT and MINOS muon and radon data sets were determined by performing a $\chi^{2}$ fit test of Eq.~(\ref{eq:cosine}) to the data described in Sec.~\ref{sec:TheData} and shown in Fig.~\ref{fig:AllData}. The results of these fits are given in Table~\ref{tab:FitResults}. The confidence limit contours for the best fit phase and period are shown in Fig.~\ref{fig:PhasePeriodFits}. \\

Our fit to the CoGeNT data is in good agreement with the published results~\cite{Aalseth:2011wp} and disfavors the null modulation hypothesis at \unit[3.1]{$\sigma$}. The significance with which we exclude the null modulation hypothesis is defined as the square root of the difference between the  $\chi^{2}$ value of the best fit point and that of the null modulation hypothesis. This definition is different from that used in the published CoGeNT analysis and gives a slightly stronger exclusion.  The small differences between our best fit values to the CoGeNT data and the published CoGeNT best fit values may be explained by the assumption in our fits that the CoGeNT errors are uncorrelated.\\

The two apparent occurrences of sudden stratospheric warming events~\cite{Osprey:2009ni} in early 2010 and early 2011, which temporarily increased the muon rate, drive the large $\chi^{2}$ for the muon fit and cause the best fit period to be significantly smaller than one year.  If the complete MINOS muon data set, August 2003 to April 2012, is fit, minimizing the impact of short term fluctuations, a period much closer to one year is obtained, $T$=\unit[(364.5 $\pm$ 0.3)]{days}, and the phase remains unchanged. \\

 The best fit phase differences $\delta t_{0}$ between the CoGeNT phase and the MINOS muon and radon phases are determined by minimizing:

\begin{footnotesize}
\begin{eqnarray}
\label{eq:LongFit}
\chi^{2}(\delta t_{0})&=&\sum_{i=1}^{N_{M}} \frac{(R_{ob,M,i}-R_{ex}(R_{0,M},A_{M},t_{0},T))^{2}}{\sigma_{M,i}^{2}}\\
       &+&\sum_{i=1}^{N_{C}=16} \frac{(R_{ob,C,i}-R_{ex}(R_{0,C},A_{C},t_{0}+\delta t_{0},T))^{2}}{\sigma_{C,i}^{2}}.\nonumber
\end{eqnarray}
\end{footnotesize}
The first term in Eq.~(\ref{eq:LongFit}) is the $\chi^{2}$ contribution from the MINOS muon rate or radon level data where $N_{M}$ is the number of live days concurrent with the CoGeNT data collection period.  The second term is the contribution from the CoGeNT event rate data.  $R_{ob,M,i}$~($R_{ob,C,i}$) is the $i^{th}$ observed MINOS~(CoGeNT) data point.  $\sigma_{M,i}$ and $\sigma_{C,j}$ are the uncertainties on the MINOS and CoGeNT data points respectively.  $R_{ex}$ is the expected value, as determined by Eq.~(\ref{eq:cosine}), assuming the given modulation parameters and  $\delta t_{0}$ is defined as the phase of the CoGeNT data minus the phase of the MINOS data.  The $\chi^{2}$, as a function of this phase difference, is determined by minimizing the $\chi^{2}$ over the MINOS mean value $R_{0,M}$, the amplitude $A_{M}$ and phase $t_{0}$ and the CoGeNT mean value $R_{0,C}$, amplitude $A_{C}$; and, for some fits, a common period $T$.\\

\begin{figure}[htb]
  \begin{center}
    \includegraphics[width=0.5\textwidth]{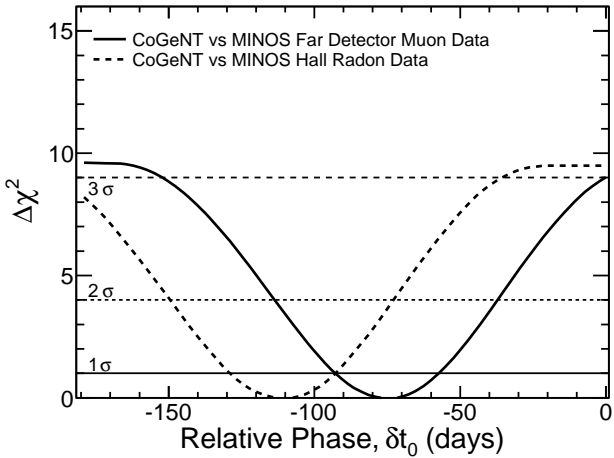}
  \end{center}
  \caption{The $\Delta\chi^{2}$ distributions comparing the phases of the MINOS muon rate and radon level data to the phase of the CoGeNT event rate data using Eq.~(\ref{eq:LongFit}).  The $\Delta\chi^{2}$ curves are calculated with respect to their $\chi^{2}$ minima. The flattening of the $\Delta\chi^{2}$ curves indicate that these exclusions are limited by the confidence with which the CoGeNT data can exclude the null modulation hypothesis. }
  \label{fig:DeltaPhaseFits}
\end{figure}

Figure \ref{fig:DeltaPhaseFits} shows the $\Delta\chi^{2}$ curves, as a function of $\delta t_{0}$, for the simultaneous fits of the MINOS and CoGeNT data to Eq.~(\ref{eq:LongFit}) assuming a common period of one year.  The best fit phase differences are \unit[(-75 $\pm$ 18)]{days} and \unit[(-110 $\pm$ 18)]{days} for the comparison to the muon and radon data respectively and \unit[(-67 $\pm$ 17)]{days} and \unit[(-112 $\pm$ 18)]{days} respectively when minimizing the $\chi^{2}$ over the period $T$. The statistical significance at which equivalent phases for the MINOS and CoGeNT data can be excluded is given by the square root of the $\Delta\chi^{2}$ difference between the best fit point and the value at $\delta t_{0}=0$. As can be seen from Fig.~\ref{fig:DeltaPhaseFits} the phases of the MINOS muon and radon data are inconsistent with the phase of the CoGeNT data at \unit[3.0]{$\sigma$} and \unit[3.1]{$\sigma$} respectively.\\
\subsection{Shape-Free $\chi^{2}$ Test}
\label{sec:ShapeFreeTest}
\begin{figure}[htb]
  \begin{center}
    \includegraphics[width=0.5\textwidth]{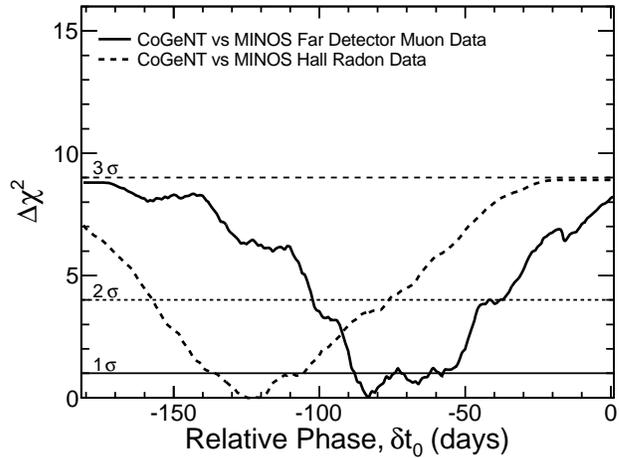}
  \end{center}
  \caption{The $\Delta\chi^{2}$ distributions comparing the phases of the MINOS muon rate and radon level data to the phase of the CoGeNT event rate data using Eq.~(\ref{eq:ShortFit}).  The $\Delta\chi^{2}$ curves are calculated with respect to their $\chi^{2}$ minima. }
  \label{fig:ShortFitPlots}
\end{figure}

In this section we determine the relative phase $\delta t_{0}$ between the MINOS and CoGeNT data sets, without an {\it a priori} assumption regarding their shape, by calculating the $\chi^{2}$ difference between their respective modulations. The $\chi^{2}$ difference, assuming a common binning, is defined as: 
\begin{equation}
\chi^{2}(\delta t_{0})= \sum_{i=1}^{N_{C}=16}\frac{(R_{C,i}-f\cdot R_{M,i}(\delta t_{0}))}{\sigma^{2}_{C,i}+\sigma^{2}_{M,i}}.
\label{eq:ShortFit}
\end{equation}
$R_{M,i}$~($R_{C,i}$) is the $i^{th}$~ MINOS~(CoGeNT) residual and $\sigma_{M,i}$ and $\sigma_{C,i}$ are the uncertainties on the MINOS and CoGeNT residuals respectively.  We marginalize over the difference in amplitudes, for each $\delta t_{0}$, by minimizing the $\chi^{2}$ over a positive definite multiplicative factor $f$.  If the data have similar underlying forms, we expect the $\chi^{2}$ to be a minimum when the phase difference between them is zero. The $\chi^{2}$ values, as a function of $\delta t$, are determined by shifting the time-axis of the MINOS data by $\delta t$ days and recalculating Eq.~(\ref{eq:ShortFit}). Figure~\ref{fig:ShortFitPlots} shows the $\Delta\chi^{2}$ curves as a function of the MINOS data offset, which is equivalent to the relative phase $\delta t_{0}$.  The curves are not smooth due to statistical fluctuations in the data. By offsetting the MINOS data we vary the number of MINOS live days which overlap the CoGeNT data.  To ensure that each subset of MINOS data, for every $\delta t$, contains the same number of live days we substitute the historical daily average of that date for those days which do not pass the live-time selection criteria.  The best fit phase differences between the CoGeNT data and the MINOS muon~(radon) data, corresponding to the minimum of the $\Delta\chi^{2}$ curves in Fig.~\ref{fig:ShortFitPlots}, are \unit[$-83^{+25}_{-5}$]{days}~(\unit[$-123^{+18}_{-16}$]{days}).  The statistical significance, as defined in Sec.~\ref{sec:CosineTest}, at which equivalent phases for the CoGeNT and MINOS muon~(radon) data are excluded is \unit[2.9]{$\sigma$}~(\unit[3.0]{$\sigma$}).  \\
\subsection{Correlation Test}
\label{sec:CorrelationTest}
Residual muon or radon backgrounds in the CoGeNT data could cause a correlation between the CoGeNT modulation and the MINOS muon and radon modulation measurements.  The degree of correlation has been evaluated using Pearson's coefficient of correlation, calculated as:
\begin{small}
\begin{equation}
\rho=\frac{1}{N_{C}-1}\sum_{i=1}^{N_{C}=16}\frac{(R_{ob,M,i}-\overline{R_{ob,M}})(R_{ob,C,i}-\overline{R_{ob,C}})}{\sigma_{M}\sigma_{C}},
\label{eq:correlation}
\end{equation}
\end{small}
where $N_{C}$ is the number of bins, $\overline{R_{ob,M}}$ and $\overline{R_{ob,C}}$ are the average values of the MINOS and CoGeNT data sets. $\sigma_{M}$ and $\sigma_{C}$ are the standard deviations of the points comprising the MINOS and CoGeNT data sets respectively. The correlation coefficients, and their Fisher transforms~\cite{Fisher:1936et}, are given in Table~\ref{tab:Correlations}.  \\
\begin{table}[htb]
\begin{tabular}{ccc} \hline
Data Set & Correlation  & Fisher\\
         & Coefficient~($\rho$)  & Transform\\ \hline
CoGeNT vs Muon Data & 0.19   & 0.19 $\pm$ 0.28 \\
CoGeNT vs Radon Data & -0.29 & -0.30 $\pm $0.28\\ \hline
\end{tabular} 
\caption{The coefficients of correlation, and their Fisher transforms, calculated between the CoGeNT event rate data and the MINOS muon rate and radon level data. Both data sets being compared are consistent with no correlation at $\sim$1$\sigma$. }
\label{tab:Correlations}
\end{table}

Even if there is no causal relationship between the observed MINOS muon and radon modulations and the CoGeNT modulation, there will be some correlation between these data sets as they all follow an approximate sinusoidal variation.  The expected value of the correlation is related to their relative phases.  For example, if the phase difference between two periodic data sets is smaller~(larger) than one-quarter of the period, the correlation should be positive~(negative).  One can therefore infer from the results in Table~\ref{tab:Correlations} that the effective phase difference between CoGeNT and the MINOS muon data is near to but less than \unit[365.25/4]{days}, while between CoGeNT and the MINOS radon data it is near to but more than  \unit[365.25/4]{days}. \\

 To verify whether the calculated correlations are consistent with the observed modulation phases we generated a series of pseudo-experiments. Sampling from two cosine curves, with the precision and binning of the CoGeNT and MINOS data sets and amplitudes taken from Table~\ref{tab:FitResults}, we calculated the Fisher transform as a function of the phase difference between the two curves. We find that the observed values of the Fisher transforms in Table~\ref{tab:Correlations}, (0.19 $\pm$ 0.28) and (-0.30 $\pm$ 0.28), correspond to phase differences of \unit[-77$^{+31}_{-47}$]{days} and \unit[-117$^{+53}_{-37}$]{days} respectively.  These values are consistent with the phase differences calculated in the preceding sections.\\
\section{Conclusion}
\label{sec:Conclusion}
We have performed a comparison of the modulation phases observed in the CoGeNT and MINOS atmospheric muon and radon data, all collected concurrently between  Dec.~4, 2009 and March~6, 2011 in the Soudan Underground Laboratory. We have presented the results of a shape-free data-to-data comparison which indicate that the phases of the CoGeNT data and the atmospheric muon and radon data are different by \unit[$-83^{+25}_{-5}$]{days}~(\unit[2.9]{$\sigma$})  and \unit[$-123^{+18}_{-16}$]{days}~(\unit[3.0]{$\sigma$}) respectively. The calculated correlation coefficients between the CoGeNT and MINOS data sets are statistically consistent with the no-correlation hypothesis. The cosine fit test measures the phase difference between the CoGeNT and MINOS muon data sets to be \unit[(-75 $\pm$ 18)]{days}, inconsistent at \unit[3.0]{$\sigma$}, and between the CoGeNT and MINOS radon data sets to be \unit[(-110 $\pm$ 18)]{days}, inconsistent at \unit[3.1]{$\sigma$}. The similarity between the results of both these tests indicate that no significant bias is introduced when imposing a sinusoidal shape on the data.  It is also clear that our exclusions are limited by the degree to which the CoGeNT data exclude the null modulation hypothesis.  Based on the studies described above, it appears unlikely that muon or radon related processes contribute significantly to the observed CoGeNT modulation.\\
\section{Acknowledgments}
\label{sec:Acknowledgements} 
This work was supported by the US DOE, the UK STFC, the US NSF, the State and University of Minnesota, the University of Athens, Greece and Brazil's FAPESP and CNPq.  We are grateful to the Minnesota Department of Natural Resources, the crew of Soudan Underground Laboratory, and the staff of Fermilab for their contributions to this effort. We also thank Juan Collar and the CoGeNT collaboration for sharing their data thus facilitating this analysis.

\bibliography{DarkMatter}

\providecommand{\noopsort}[1]{}\providecommand{\singleletter}[1]{#1}%
\begin{thebibliography}{49}
\expandafter\ifx\csname natexlab\endcsname\relax\def\natexlab#1{#1}\fi
\expandafter\ifx\csname bibnamefont\endcsname\relax
  \def\bibnamefont#1{#1}\fi
\expandafter\ifx\csname bibfnamefont\endcsname\relax
  \def\bibfnamefont#1{#1}\fi
\expandafter\ifx\csname citenamefont\endcsname\relax
  \def\citenamefont#1{#1}\fi
\expandafter\ifx\csname url\endcsname\relax
  \def\url#1{\texttt{#1}}\fi
\expandafter\ifx\csname urlprefix\endcsname\relax\def\urlprefix{URL }\fi
\providecommand{\bibinfo}[2]{#2}
\providecommand{\eprint}[2][]{\url{#2}}

\bibitem[{\citenamefont{Feng}(2010)}]{Feng:2010gw}
\bibinfo{author}{\bibfnamefont{J.~L.} \bibnamefont{Feng}},
  \bibinfo{journal}{Ann. Rev. Astron. Astrophys.}
  \textbf{\bibinfo{volume}{48}}, \bibinfo{pages}{495} (\bibinfo{year}{2010}),
  \eprint{astro-ph/1003.0904}.

\bibitem[{\citenamefont{Bertone et~al.}(2005)\citenamefont{Bertone, Hooper, and
  Silk}}]{Bertone:2004pz}
\bibinfo{author}{\bibfnamefont{G.}~\bibnamefont{Bertone}},
  \bibinfo{author}{\bibfnamefont{D.}~\bibnamefont{Hooper}}, \bibnamefont{and}
  \bibinfo{author}{\bibfnamefont{J.}~\bibnamefont{Silk}},
  \bibinfo{journal}{Phys. Rept.} \textbf{\bibinfo{volume}{405}},
  \bibinfo{pages}{279} (\bibinfo{year}{2005}), \eprint{hep-ph/0404175}.

\bibitem[{\citenamefont{Drukier et~al.}(1986)\citenamefont{Drukier, Freese, and
  Spergel}}]{Drukier:1986tm}
\bibinfo{author}{\bibfnamefont{A.~K.} \bibnamefont{Drukier}},
  \bibinfo{author}{\bibfnamefont{K.}~\bibnamefont{Freese}}, \bibnamefont{and}
  \bibinfo{author}{\bibfnamefont{D.~N.} \bibnamefont{Spergel}},
  \bibinfo{journal}{Phys. Rev. D} \textbf{\bibinfo{volume}{33}},
  \bibinfo{pages}{3495} (\bibinfo{year}{1986}).

\bibitem[{\citenamefont{Freese et~al.}(1988)\citenamefont{Freese, Frieman, and
  Gould}}]{Freese:1987wu}
\bibinfo{author}{\bibfnamefont{K.}~\bibnamefont{Freese}},
  \bibinfo{author}{\bibfnamefont{J.~A.} \bibnamefont{Frieman}},
  \bibnamefont{and} \bibinfo{author}{\bibfnamefont{A.}~\bibnamefont{Gould}},
  \bibinfo{journal}{Phys. Rev. D} \textbf{\bibinfo{volume}{37}},
  \bibinfo{pages}{3388} (\bibinfo{year}{1988}).

\bibitem[{\citenamefont{Frandsen et~al.}(2012)}]{Frandsen:2011gi}
\bibinfo{author}{\bibfnamefont{M.~T.} \bibnamefont{Frandsen}}
  \bibnamefont{et~al.}, \bibinfo{journal}{J. Cosm. Astropart. Phys.}
  \textbf{\bibinfo{volume}{1201}}, \bibinfo{pages}{024} (\bibinfo{year}{2012}),
  \eprint{hep-ph/1111.0292}.

\bibitem[{\citenamefont{Aalseth et~al.}(2011{\natexlab{a}})}]{Aalseth:2011wp}
\bibinfo{author}{\bibfnamefont{C.~E.} \bibnamefont{Aalseth}}
  \bibnamefont{et~al.} (\bibinfo{collaboration}{CoGeNT Collaboration}),
  \bibinfo{journal}{Phys. Rev. Lett.} \textbf{\bibinfo{volume}{107}},
  \bibinfo{pages}{141301} (\bibinfo{year}{2011}{\natexlab{a}}),
  \eprint{astro-ph/1106.0650}.

\bibitem[{\citenamefont{Aalseth et~al.}(2011{\natexlab{b}})}]{Aalseth:2010vx}
\bibinfo{author}{\bibfnamefont{C.~E.} \bibnamefont{Aalseth}}
  \bibnamefont{et~al.} (\bibinfo{collaboration}{CoGeNT Collaboration}),
  \bibinfo{journal}{Phys. Rev. Lett.} \textbf{\bibinfo{volume}{106}},
  \bibinfo{pages}{131301} (\bibinfo{year}{2011}{\natexlab{b}}),
  \eprint{astro-ph/1002.4703}.

\bibitem[{\citenamefont{Bernabei et~al.}(2010)}]{Bernabei:2010mq}
\bibinfo{author}{\bibfnamefont{R.}~\bibnamefont{Bernabei}}
  \bibnamefont{et~al.}, \bibinfo{journal}{Eur. Phys. J. C}
  \textbf{\bibinfo{volume}{67}}, \bibinfo{pages}{39} (\bibinfo{year}{2010}),
  \eprint{astro-ph/1002.1028}.

\bibitem[{\citenamefont{Angloher et~al.}(2012)}]{Angloher:2011uu}
\bibinfo{author}{\bibfnamefont{G.}~\bibnamefont{Angloher}} \bibnamefont{et~al.}
  (\bibinfo{collaboration}{CRESST II Collaboration}), \bibinfo{journal}{Eur.
  Phys. J. C} \textbf{\bibinfo{volume}{72}}, \bibinfo{pages}{1971}
  (\bibinfo{year}{2012}), \eprint{astro-ph/1109.0702}.

\bibitem[{\citenamefont{Bernabei et~al.}(2008)}]{Bernabei:2008yh}
\bibinfo{author}{\bibfnamefont{R.}~\bibnamefont{Bernabei}} \bibnamefont{et~al.}
  (\bibinfo{collaboration}{DAMA Collaboration}), \bibinfo{journal}{Nucl.
  Instrum. Meth. A} \textbf{\bibinfo{volume}{592}}, \bibinfo{pages}{297}
  (\bibinfo{year}{2008}), \eprint{astro-ph/0804.2738}.

\bibitem[{\citenamefont{Hooper}(2012)}]{Hooper:2012ft}
\bibinfo{author}{\bibfnamefont{D.}~\bibnamefont{Hooper}}
  (\bibinfo{year}{2012}), \eprint{astro-ph/1201.1303}.

\bibitem[{\citenamefont{Kelso et~al.}(2012)\citenamefont{Kelso, Hooper, and
  Buckley}}]{Kelso:2011gd}
\bibinfo{author}{\bibfnamefont{C.}~\bibnamefont{Kelso}},
  \bibinfo{author}{\bibfnamefont{D.}~\bibnamefont{Hooper}}, \bibnamefont{and}
  \bibinfo{author}{\bibfnamefont{M.~R.} \bibnamefont{Buckley}},
  \bibinfo{journal}{Phys. Rev. D} \textbf{\bibinfo{volume}{85}},
  \bibinfo{pages}{043515} (\bibinfo{year}{2012}), \eprint{astro-ph/1110.5338}.

\bibitem[{\citenamefont{Kopp et~al.}(2012)\citenamefont{Kopp, Schwetz, and
  Zupan}}]{Kopp:2011yr}
\bibinfo{author}{\bibfnamefont{J.}~\bibnamefont{Kopp}},
  \bibinfo{author}{\bibfnamefont{T.}~\bibnamefont{Schwetz}}, \bibnamefont{and}
  \bibinfo{author}{\bibfnamefont{J.}~\bibnamefont{Zupan}}, \bibinfo{journal}{J.
  Cosm. Astropart. Phys.} \textbf{\bibinfo{volume}{1203}}, \bibinfo{pages}{001}
  (\bibinfo{year}{2012}), \eprint{hep-ph/1110.2721}.

\bibitem[{\citenamefont{Ahmed et~al.}(2011{\natexlab{a}})}]{Ahmed:2010hw}
\bibinfo{author}{\bibfnamefont{Z.}~\bibnamefont{Ahmed}} \bibnamefont{et~al.}
  (\bibinfo{collaboration}{CDMS-II Collaboration}), \bibinfo{journal}{Phys.
  Rev. D} \textbf{\bibinfo{volume}{83}}, \bibinfo{pages}{112002}
  (\bibinfo{year}{2011}{\natexlab{a}}), \eprint{astro-ph/1012.5078}.

\bibitem[{\citenamefont{Ahmed et~al.}(2011{\natexlab{b}})}]{Ahmed:2010wy}
\bibinfo{author}{\bibfnamefont{Z.}~\bibnamefont{Ahmed}} \bibnamefont{et~al.}
  (\bibinfo{collaboration}{CDMS-II Collaboration}), \bibinfo{journal}{Phys.
  Rev. Lett.} \textbf{\bibinfo{volume}{106}}, \bibinfo{pages}{131302}
  (\bibinfo{year}{2011}{\natexlab{b}}), \eprint{astro-ph/1011.2482}.

\bibitem[{\citenamefont{Ahmed et~al.}(2012)}]{Ahmed:2012vq}
\bibinfo{author}{\bibfnamefont{Z.}~\bibnamefont{Ahmed}} \bibnamefont{et~al.}
  (\bibinfo{collaboration}{CDMS Collaboration}) (\bibinfo{year}{2012}),
  \eprint{astro-ph/1203.1309}.

\bibitem[{\citenamefont{Angle et~al.}(2011)}]{Angle:2011th}
\bibinfo{author}{\bibfnamefont{J.}~\bibnamefont{Angle}} \bibnamefont{et~al.}
  (\bibinfo{collaboration}{XENON10 Collaboration}), \bibinfo{journal}{Phys.
  Rev. Lett.} \textbf{\bibinfo{volume}{107}}, \bibinfo{pages}{051301}
  (\bibinfo{year}{2011}), \eprint{astro-ph/1104.3088}.

\bibitem[{\citenamefont{Aprile et~al.}(2011)}]{Aprile:2011hi}
\bibinfo{author}{\bibfnamefont{E.}~\bibnamefont{Aprile}} \bibnamefont{et~al.}
  (\bibinfo{collaboration}{XENON100 Collaboration}), \bibinfo{journal}{Phys.
  Rev. Lett.} \textbf{\bibinfo{volume}{107}}, \bibinfo{pages}{131302}
  (\bibinfo{year}{2011}), \eprint{astro-ph/1104.2549}.

\bibitem[{\citenamefont{Aprile et~al.}(2012)}]{Aprile:2012hi}
\bibinfo{author}{\bibfnamefont{E.}~\bibnamefont{Aprile}} \bibnamefont{et~al.}
  (\bibinfo{collaboration}{XENON100 Collaboration}), \bibinfo{journal}{Phys.
  Rev. Lett.} \textbf{\bibinfo{volume}{109}}, \bibinfo{pages}{181301}
  (\bibinfo{year}{2012}), \eprint{astro-ph/1207.5988}.

\bibitem[{\citenamefont{Armengaud et~al.}(2012)}]{Armengaud:2012kd}
\bibinfo{author}{\bibfnamefont{E.}~\bibnamefont{Armengaud}}
  \bibnamefont{et~al.} (\bibinfo{collaboration}{EDELWEISS Collaboration}),
  \bibinfo{journal}{Phys. Rev. D} \textbf{\bibinfo{volume}{86}},
  \bibinfo{pages}{051701} (\bibinfo{year}{2012}), \eprint{astro-ph/1207.1815}.

\bibitem[{\citenamefont{Herrero-Garcia
  et~al.}(2012{\natexlab{a}})\citenamefont{Herrero-Garcia, Schwetz, and
  Zupan}}]{HerreroGarcia:2012fu}
\bibinfo{author}{\bibfnamefont{J.}~\bibnamefont{Herrero-Garcia}},
  \bibinfo{author}{\bibfnamefont{T.}~\bibnamefont{Schwetz}}, \bibnamefont{and}
  \bibinfo{author}{\bibfnamefont{J.}~\bibnamefont{Zupan}},
  \bibinfo{journal}{Phys. Rev. Lett.} \textbf{\bibinfo{volume}{109}},
  \bibinfo{pages}{141301} (\bibinfo{year}{2012}{\natexlab{a}}),
  \eprint{hep-ph/1205.0134}.

\bibitem[{\citenamefont{Collar and Fields}(2012)}]{Collar:2012ed}
\bibinfo{author}{\bibfnamefont{J.}~\bibnamefont{Collar}} \bibnamefont{and}
  \bibinfo{author}{\bibfnamefont{N.}~\bibnamefont{Fields}}
  (\bibinfo{year}{2012}), \eprint{astro-ph/1204.3559}.

\bibitem[{\citenamefont{Collar}(2011)}]{Collar:2011wq}
\bibinfo{author}{\bibfnamefont{J.}~\bibnamefont{Collar}}
  (\bibinfo{year}{2011}), \eprint{astro-ph/1106.0653}.

\bibitem[{\citenamefont{Schwetz and Zupan}(2011)}]{Schwetz:2011xm}
\bibinfo{author}{\bibfnamefont{T.}~\bibnamefont{Schwetz}} \bibnamefont{and}
  \bibinfo{author}{\bibfnamefont{J.}~\bibnamefont{Zupan}}, \bibinfo{journal}{J.
  Cosm. Astropart. Phys.} \textbf{\bibinfo{volume}{1108}}, \bibinfo{pages}{008}
  (\bibinfo{year}{2011}), \eprint{hep-ph/1106.6241}.

\bibitem[{\citenamefont{Farina et~al.}(2011)\citenamefont{Farina, Pappadopulo,
  Strumia, and Volansky}}]{Farina:2011pw}
\bibinfo{author}{\bibfnamefont{M.}~\bibnamefont{Farina}},
  \bibinfo{author}{\bibfnamefont{D.}~\bibnamefont{Pappadopulo}},
  \bibinfo{author}{\bibfnamefont{A.}~\bibnamefont{Strumia}}, \bibnamefont{and}
  \bibinfo{author}{\bibfnamefont{T.}~\bibnamefont{Volansky}},
  \bibinfo{journal}{J. Cosm. Astropart. Phys.} \textbf{\bibinfo{volume}{1111}},
  \bibinfo{pages}{010} (\bibinfo{year}{2011}), \eprint{hep-ph/1107.0715}.

\bibitem[{\citenamefont{Foot}(2010)}]{Foot:2010rj}
\bibinfo{author}{\bibfnamefont{R.}~\bibnamefont{Foot}}, \bibinfo{journal}{Phys.
  Lett. B} \textbf{\bibinfo{volume}{692}}, \bibinfo{pages}{65}
  (\bibinfo{year}{2010}), \eprint{hep-ph/1004.1424}.

\bibitem[{\citenamefont{Herrero-Garcia
  et~al.}(2012{\natexlab{b}})\citenamefont{Herrero-Garcia, Schwetz, and
  Zupan}}]{HerreroGarcia:2011aa}
\bibinfo{author}{\bibfnamefont{J.}~\bibnamefont{Herrero-Garcia}},
  \bibinfo{author}{\bibfnamefont{T.}~\bibnamefont{Schwetz}}, \bibnamefont{and}
  \bibinfo{author}{\bibfnamefont{J.}~\bibnamefont{Zupan}}, \bibinfo{journal}{J.
  Cosm. Astropart. Phys.} \textbf{\bibinfo{volume}{1203}}, \bibinfo{pages}{005}
  (\bibinfo{year}{2012}{\natexlab{b}}), \eprint{hep-ph/1112.1627}.

\bibitem[{\citenamefont{Ralston}(2010)}]{Ralston:2010bd}
\bibinfo{author}{\bibfnamefont{J.~P.} \bibnamefont{Ralston}}
  (\bibinfo{year}{2010}), \eprint{hep-ph/1006.5255}.

\bibitem[{\citenamefont{Blum}(2011)}]{Blum:2011jf}
\bibinfo{author}{\bibfnamefont{K.}~\bibnamefont{Blum}} (\bibinfo{year}{2011}),
  \eprint{astro-ph/1110.0857}.

\bibitem[{\citenamefont{Aalseth et~al.}(2012)}]{Aalseth:2012if}
\bibinfo{author}{\bibfnamefont{C.}~\bibnamefont{Aalseth}} \bibnamefont{et~al.}
  (\bibinfo{collaboration}{CoGeNT Collaboration}) (\bibinfo{year}{2012}),
  \eprint{astro-ph/1208.5737}.

\bibitem[{\citenamefont{Ambrosio et~al.}(1997)}]{Ambrosio:1997tc}
\bibinfo{author}{\bibfnamefont{M.}~\bibnamefont{Ambrosio}} \bibnamefont{et~al.}
  (\bibinfo{collaboration}{MACRO Collaboration}), \bibinfo{journal}{Astropart.
  Phys.} \textbf{\bibinfo{volume}{7}}, \bibinfo{pages}{109}
  (\bibinfo{year}{1997}).

\bibitem[{\citenamefont{Bouchta}(1999)}]{Bouchta:1999}
\bibinfo{author}{\bibfnamefont{A.}~\bibnamefont{Bouchta}}
  (\bibinfo{collaboration}{AMANDA Collaboration}),
  \bibinfo{howpublished}{{Proceedings of the 26th ICRC}, vol. 2, pp. 108-111}
  (\bibinfo{year}{1999}).

\bibitem[{\citenamefont{Bellini et~al.}(2012)}]{Bellini:2012te}
\bibinfo{author}{\bibfnamefont{G.}~\bibnamefont{Bellini}} \bibnamefont{et~al.}
  (\bibinfo{collaboration}{Borexino Collaboration}), \bibinfo{journal}{J. Cosm.
  Astropart. Phys.} \textbf{\bibinfo{volume}{1205}}, \bibinfo{pages}{015}
  (\bibinfo{year}{2012}), \eprint{hep-ex/1202.6403}.

\bibitem[{\citenamefont{Selvi}(2009)}]{Solvi:2009}
\bibinfo{author}{\bibfnamefont{M.}~\bibnamefont{Selvi}}
  (\bibinfo{collaboration}{LVD Collaboration}),
  \bibinfo{howpublished}{{Proceedings of the 31st ICRC}}
  (\bibinfo{year}{2009}).

\bibitem[{\citenamefont{Desiati et~al.}(2011)}]{Desiati:2011}
\bibinfo{author}{\bibfnamefont{P.}~\bibnamefont{Desiati}} \bibnamefont{et~al.}
  (\bibinfo{collaboration}{ICECUBE Collaboration}),
  \bibinfo{howpublished}{{Proceedings of the 32nd ICRC}}
  (\bibinfo{year}{2011}), \eprint{astro-ph/1111.2735}.

\bibitem[{\citenamefont{Adamson et~al.}(2010)}]{Adamson:2009zf}
\bibinfo{author}{\bibfnamefont{P.}~\bibnamefont{Adamson}} \bibnamefont{et~al.}
  (\bibinfo{collaboration}{MINOS Collaboration}), \bibinfo{journal}{Phys. Rev.
  D} \textbf{\bibinfo{volume}{81}}, \bibinfo{pages}{012001}
  (\bibinfo{year}{2010}), \eprint{hep-ex/0909.4012}.

\bibitem[{\citenamefont{Bernabei et~al.}(2012)}]{Bernabei:2012wp}
\bibinfo{author}{\bibfnamefont{R.}~\bibnamefont{Bernabei}}
  \bibnamefont{et~al.}, \bibinfo{journal}{Eur. Phys. J. C}
  \textbf{\bibinfo{volume}{72}}, \bibinfo{pages}{2064} (\bibinfo{year}{2012}),
  \eprint{astro-ph/1202.4179}.

\bibitem[{\citenamefont{Fernandez-Martinez and
  Mahbubani}(2012)}]{FernandezMartinez:2012wd}
\bibinfo{author}{\bibfnamefont{E.}~\bibnamefont{Fernandez-Martinez}}
  \bibnamefont{and}
  \bibinfo{author}{\bibfnamefont{R.}~\bibnamefont{Mahbubani}},
  \bibinfo{journal}{J. Cosm. Astropart. Phys.} \textbf{\bibinfo{volume}{1207}},
  \bibinfo{pages}{029} (\bibinfo{year}{2012}), \eprint{astro-ph/1204.5180}.

\bibitem[{\citenamefont{Chang et~al.}(2012)\citenamefont{Chang, Pradler, and
  Yavin}}]{Chang:2011eb}
\bibinfo{author}{\bibfnamefont{S.}~\bibnamefont{Chang}},
  \bibinfo{author}{\bibfnamefont{J.}~\bibnamefont{Pradler}}, \bibnamefont{and}
  \bibinfo{author}{\bibfnamefont{I.}~\bibnamefont{Yavin}},
  \bibinfo{journal}{Phys. Rev. D} \textbf{\bibinfo{volume}{85}},
  \bibinfo{pages}{063505} (\bibinfo{year}{2012}), \eprint{hep-ph/1111.4222}.

\bibitem[{\citenamefont{Goodman}(1999)}]{Goodman:1999}
\bibinfo{author}{\bibfnamefont{M.}~\bibnamefont{Goodman}}
  (\bibinfo{collaboration}{Soudan 2 Collaboration}),
  \bibinfo{howpublished}{{Proceedings of the 26th ICRC}, vol. 2, pp. 324-327}
  (\bibinfo{year}{1999}).

\bibitem[{\citenamefont{Barbeau et~al.}(2007)\citenamefont{Barbeau, Collar, and
  Tench}}]{Barbeau:2007qi}
\bibinfo{author}{\bibfnamefont{P.}~\bibnamefont{Barbeau}},
  \bibinfo{author}{\bibfnamefont{J.}~\bibnamefont{Collar}}, \bibnamefont{and}
  \bibinfo{author}{\bibfnamefont{O.}~\bibnamefont{Tench}}, \bibinfo{journal}{J.
  Cosm. Astropart. Phys.} \textbf{\bibinfo{volume}{0709}}, \bibinfo{pages}{009}
  (\bibinfo{year}{2007}), \eprint{nucl-ex/0701012}.

\bibitem[{\citenamefont{Michael et~al.}(2008)}]{Michael:2008bc}
\bibinfo{author}{\bibfnamefont{D.~G.} \bibnamefont{Michael}}
  \bibnamefont{et~al.} (\bibinfo{collaboration}{MINOS Collaboration}),
  \bibinfo{journal}{Nucl. Instrum. Meth. A} \textbf{\bibinfo{volume}{596}},
  \bibinfo{pages}{190} (\bibinfo{year}{2008}),
  \eprint{physics.ins-det/0805.3170}.

\bibitem[{\citenamefont{Crane et~al.}(1995)}]{Crane:1995ky}
\bibinfo{author}{\bibfnamefont{D.~A.} \bibnamefont{Crane}} \bibnamefont{et~al.}
  (\bibinfo{collaboration}{NuMI Beam Group}) (\bibinfo{year}{1995}),
  \bibinfo{note}{{FERMILAB-TM-1946}}.

\bibitem[{\citenamefont{Aalseth et~al.}(2008)}]{Aalseth:2008rx}
\bibinfo{author}{\bibfnamefont{C.~E.} \bibnamefont{Aalseth}}
  \bibnamefont{et~al.} (\bibinfo{collaboration}{CoGeNT Collaboration}),
  \bibinfo{journal}{Phys. Rev. Lett.} \textbf{\bibinfo{volume}{101}},
  \bibinfo{pages}{251301} (\bibinfo{year}{2008}), \eprint{astro-ph/0807.0879}.

\bibitem[{\citenamefont{Adamson et~al.}(2007)}]{Adamson:2007ww}
\bibinfo{author}{\bibfnamefont{P.}~\bibnamefont{Adamson}} \bibnamefont{et~al.}
  (\bibinfo{collaboration}{MINOS}), \bibinfo{journal}{Phys. Rev. D}
  \textbf{\bibinfo{volume}{76}}, \bibinfo{pages}{052003}
  (\bibinfo{year}{2007}), \eprint{hep-ex/0705.3815}.

\bibitem[{\citenamefont{{Sun Nuclear Corporation}}(2009)}]{radon-monitor}
\bibinfo{author}{\bibnamefont{{Sun Nuclear Corporation}}}
  (\bibinfo{year}{2009}), \bibinfo{note}{{1027 Radon Monitor User's Guide}},
  \urlprefix\url{radon.sunnuclear.com/1027/1027.asp}.

\bibitem[{\citenamefont{Freese et~al.}(2012)\citenamefont{Freese, Lisanti, and
  Savage}}]{Freese:2012xd}
\bibinfo{author}{\bibfnamefont{K.}~\bibnamefont{Freese}},
  \bibinfo{author}{\bibfnamefont{M.}~\bibnamefont{Lisanti}}, \bibnamefont{and}
  \bibinfo{author}{\bibfnamefont{C.}~\bibnamefont{Savage}}
  (\bibinfo{year}{2012}), \eprint{astro-ph/1209.3339}.

\bibitem[{\citenamefont{Osprey et~al.}(2009)}]{Osprey:2009ni}
\bibinfo{author}{\bibfnamefont{S.}~\bibnamefont{Osprey}} \bibnamefont{et~al.}
  (\bibinfo{collaboration}{MINOS Collaboration}), \bibinfo{journal}{Geophys.
  Res. Lett.} \textbf{\bibinfo{volume}{36}}, \bibinfo{pages}{L05809}
  (\bibinfo{year}{2009}).

\bibitem[{\citenamefont{Fisher}(1936)}]{Fisher:1936et}
\bibinfo{author}{\bibfnamefont{R.~A.} \bibnamefont{Fisher}},
  \bibinfo{journal}{Annals Eugen.} \textbf{\bibinfo{volume}{7}},
  \bibinfo{pages}{179} (\bibinfo{year}{1936}).

\end{thebibliography}

\end{document}